# THE LARGE MOMENTUM TRANSFER BEHAVIOUR OF MESON-PHOTON TRANSITION FORM FACTORS

PETER KROLL[1]

*Fachbereich Physik, Universität Wuppertal,*
*D-42097 Wuppertal, Germany*

ABSTRACT

It is reported on predictions for the $\pi$-$\gamma$ transition form factor obtained within a perturbative approach which includes transverse momentum effects and Sudakov corrections. The results clearly favor distribution amplitudes close to the asymptotic form, $\sim x_1 x_2$, and disfavor distribution amplitudes which are strongly concentrated in the end-point regions. Applications of that approach to the $\eta$-$\gamma$ and $\eta'$-$\gamma$ transition form factors are discussed as well.

## 1. Introduction

Hadronic form factors at large momentum transfer, $Q$, provide information on the constituents the hadrons are built up and on the dynamics controlling their interactions. Therefore, the form factors always found much interest and many papers, both theoretical and experimental ones, are devoted to them. Recently a new perturbative approach has been proposed by Botts, Li and Sterman [1,2] which allows to calculate the large $Q$ behavior of form factors. In this new approach, which one may term the modified hard scattering approach (HSA), the transverse degrees of freedom as well as Sudakov suppressions are taken into account in contrast to the standard perturbative approach [3]. The simplest cases to apply the modified HSA are the pseudoscalar meson-photon transition form factors for which data in the few GeV region is now available [5,6,7]. I am going to report on an analysis of these form factors carried out by Jakob, Raulfs and myself [4]. These transition form factors are exceptional cases in so far as, to lowest order, they are QED processes; QCD only provides corrections of the order of $10 - 20\%$ and higher Fock state contributions are suppressed by powers of $\alpha_s/Q^2$ (see also the discussion in Ref.[8]). For these reasons one may expect the modified HSA to be applicable for $Q$ larger than about 1 GeV. Input to calculations within the modified HSA are the hadronic wave functions which contain the long-distance physics and are not calculable at present. However, since the pion wave function is rather well constrained, a reliable estimate of the $\pi$-$\gamma$ form factor can be made. A corresponding calculation of the $\eta$-$\gamma$ and $\eta'$-$\gamma$ form factors allows to determine the decay constants and the mixing angle for pseudoscalar mesons.

---

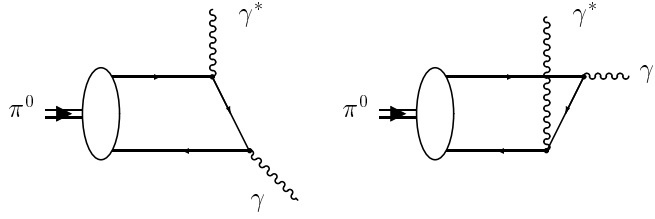

Figure 1: The basic graphs for the meson-photon transition form factor.

## 2. The $\pi$-$\gamma$ transition form factor

Adapting the modified HSA to the case of $\pi$-$\gamma$ transitions we write the corresponding form factor as

$$F_{\pi\gamma}(Q^2) = \int dx_1 \frac{d^2 b}{4\pi}\, \hat{\Psi}_0(x_1, -\mathbf{b})\, \hat{T}_H(x_1, \mathbf{b}, Q)\, \exp\left[-S(x_1, b, Q)\right], \qquad (1)$$

where $\mathbf{b}$ is the quark-antiquark separation in the transverse configuration space. $x_1$ and $x_2 = 1 - x_1$ denote the usual momentum fractions the quark and the antiquark carry, respectively. $\hat{T}_H$, the Fourier transform of the momentum space hard scattering amplitude, is to be calculated from the Feynman graphs shown in Fig. 1. One finds

$$\hat{T}_H(x_1, \mathbf{b}, Q) = \frac{2\sqrt{6}\, C_\pi}{\pi} K_0(\sqrt{x_2} Q b), \qquad (2)$$

where the charge factor $C_\pi$ is $1/3\sqrt{2}$ and $K_0$ is the modified Bessel function of order zero. The Sudakov exponent $S$ in (1) comprising those gluonic radiative corrections not taken into account by the usual QCD evolution, is given by

$$S(x_1, Q, b) = s(x_1, Q, b) + s(x_2, Q, b) - \frac{4}{\beta_0} \ln \frac{\ln(t/\Lambda_{QCD})}{\ln(1/b\Lambda_{QCD})} \qquad (3)$$

where a Sudakov function $s$ appears for each quark line entering the hard scattering amplitude ($\beta_0 = 11 - 2/3\, n_f$; $\Lambda_{QCD} = 200\,\mathrm{MeV}$). $t$ is taken to be the largest mass scale appearing in $T_H$, i. e. $t = \max(\sqrt{x_2} Q, 1/b)$. For small $b$ there is no suppression from the Sudakov factor; as $b$ increases the Sudakov factor decreases, reaching zero at $b = 1/\Lambda_{QCD}$. For even larger $b$ the Sudakov factor is set to zero. The Sudakov factor has been calculated by Botts and Sterman [1] using resummation techniques; the explicit form of the Sudakov function $s$ can be found there.

The quantity $\hat{\Psi}_0$ appearing in (1) represents the soft part of the transverse configuration space pion wave function, i. e., the full wave function with the perturbative tail removed from it. It is written as [9]

$$\hat{\Psi}_0(x_1, \mathbf{b}) = \frac{f_\pi}{2\sqrt{6}}\, \phi(x_1)\, \hat{\Sigma}(\sqrt{x_1 x_2}\, b). \qquad (4)$$

where $f_\pi (= 130.7\,\mathrm{MeV})$ is the usual pion decay constant. The wave function does not factorize in $x_1$ and $b$, but, in accord with the basic properties of the HSA[10,11], the $b$-dependence rather appears in the combination $\sqrt{x_1 x_2}\, b$. The transverse part of the wave function is assumed to be a simple Gaussian

$$\hat{\Sigma}(\sqrt{x_1 x_2}\, b) = 4\pi\, \exp\left(-x_1 x_2\, b^2/4a^2\right). \quad (5)$$

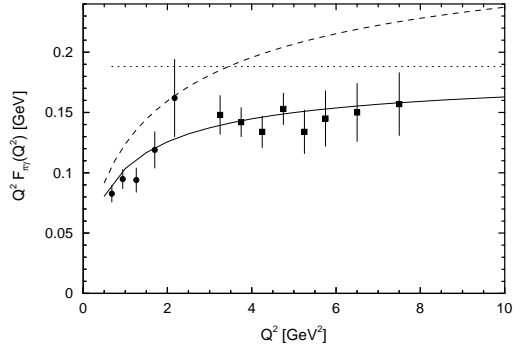

Figure 2: The $\pi$-$\gamma$ transition form factor vs. $Q^2$. The solid (dashed) line represents the prediction obtained with the modified HSA using the AS (CZ) wave function. The dotted line represents the asymptotic result. Data are taken from Ref.[7,13].

More complicated forms than (5) are proposed in Ref. [11] on the basis of dispersion relations and duality. At large transverse momentum, however, the soft momentum space wave function should behave like a Gaussian [11]. The examination of a number of examples corroborates the expectation that forms of $\hat{\Sigma}$ other than (5) will not change the results and the conclusions presented in Ref. [4] markedly.

For the distribution amplitude, $\phi$, the asymptotic form

$$\phi_{AS}(x_1) = 6\, x_1 x_2 \quad (6)$$

is used and alternatively, as a representative of strongly end-point concentrated distribution amplitudes, a form proposed by Chernyak and Zhitnitsky[12]

$$\phi_{CZ}(x_1) = 30\, x_1 x_2\, (x_1 - x_2)^2. \quad (7)$$

It is controversial whether or not (7) is supported by QCD sum rules. It is taken as an example whose significance is given by its frequent use. By QCD evolution (see, e. g. Ref. [3]) any distribution amplitude evolves into $\phi_{AS}(x)$ asymptotically, i. e. for $\ln(Q/\mu_0) \to \infty$; the asymptotic distribution amplitude itself shows no evolution. As has been shown in Ref.[10] the decay processes $\pi^+ \to \mu^+ \nu_\mu$ and $\pi^0 \to \gamma\gamma$ provide constraints on the pion wave function. Whereas the first constraint is automatically satisfied by the ansatz (4), does the second constraint fix the parameter $a$ in (5) (861 (673) MeV for the AS (CZ) wave function).

Numerical results for the transition form factor $F_{\pi\gamma}(Q^2)$ obtained from (1) are displayed in Fig. 2. It should be emphasized that there is no free parameter to be adjusted once the wave function is chosen. Obviously the results obtained from the AS wave function are in very good agreement with the CELLO data [7] and the new CLEO results presented at this conference [13]; there is not much room left for contributions from higher order perturbative QCD and/or from higher Fock states. The results obtained from the CZ wave function overshoot the data significantly. Of course, mild modifications of the asymptotic wave functions are possible without worsening the agreement between theory and experiment considerably. For example, if one follows Brodsky et al.[10] and multiplies (6) by the exponential $\exp(-a^2 m_q^2/x_1 x_2)$ where the parameter $m_q$ represents a constituent quark mass of, say, $330\,\text{MeV}$, one finds similarly good results from this modified AS wave function as from (6) itself. On the other hand, strongly end-point concentrated wave functions are clearly in conflict with the data.

The standard HSA [3] predicts for the $\pi$-$\gamma$ transition form factor

$$F_{\pi\gamma}(Q^2) = \sqrt{2}/3\, f_\pi \langle x_1^{-1} \rangle Q^{-2}. \qquad (8)$$

The bracket term denotes the $x_1^{-1}$ moment of the distribution amplitude. This moment receives the values 3 and 5 for the AS and the CZ distribution amplitude respectively. Obviously, the standard HSA, while exact at large $Q$, fails to describe the data in the few GeV region. This is to be contrasted with the modified HSA in which the QCD corrections, condensed in the Sudakov factor, and the transverse degrees of freedom provide the required $Q$-dependence. Asymptotically, the Sudakov factor damps any contribution except those from configurations with small quark-antiquark separation and, as the limiting behavior, the QCD prediction [3,14] $F_{\pi\gamma} \to \sqrt{2} f_\pi\, Q^{-2}$ emerges.

## 3. The $\eta$-$\gamma$ and $\eta'$-$\gamma$ transition form factors

The generalization of (1) to the cases of $\eta$-$\gamma$ and $\eta'$-$\gamma$ transitions. starts with the SU(3) basis states, $\eta_8$ and $\eta_1$, and the usual mixing scheme

$$|\eta\rangle = \cos\vartheta_P\, |\eta_8\rangle - \sin\vartheta_P\, |\eta_1\rangle, \qquad |\eta'\rangle = \sin\vartheta_P\, |\eta_8\rangle + \cos\vartheta_P\, |\eta_1\rangle. \qquad (9)$$

Insertion of this scheme into the $\eta$-$\gamma$ and $\eta'$-$\gamma$ matrix elements of the electromagnetic current leads to relations between the physical transition form factors and the $\eta_8$-$\gamma$ and the $\eta_1$-$\gamma$ ones. The latter form factors can be calculated analogously to the $\pi$-$\gamma$ case. The $\eta_8$ and $\eta_1$ wave functions are assumed to be identically to the AS pion wave function except of the decay constants $f_i$. In the hard scattering amplitude (2) the charge factor of the pion is to be replaced by either $C_8 = 1/3\sqrt{6}$ or $C_1 = 2/3\sqrt{3}$. Furthermore it is corrected for the rather large masses of the $\eta_i$-mesons ($m_8 = 566\,\text{MeV}$, $m_1 = 947\,\text{MeV}$, see Ref.[15]). Since the values of the decay constants and the mixing angle are not known with sufficient accuracy one can not predict the form factors but, encouraged by the success of the modified HSA in the $\pi$-$\gamma$ case, rather try to determine these parameters. Admittedly, additional information is required for this task since the $\eta$-$\gamma$ and $\eta'$-$\gamma$ transition form factors do not suffice to fix the three parameters; for any value of the mixing angle a reasonable fit to the data is obtained. The necessary extra information is provided by the two-photon decays

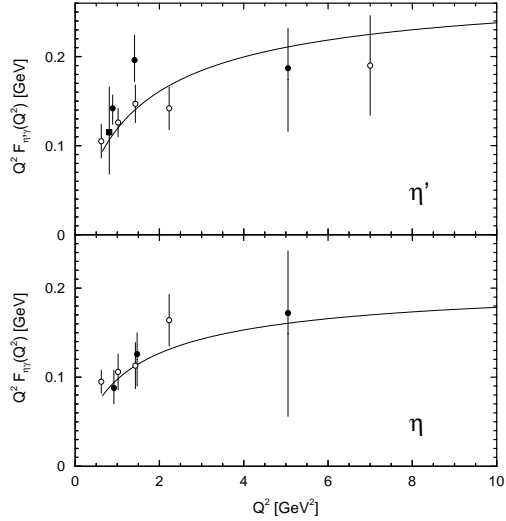

Figure 3: The $\eta$-$\gamma$ and $\eta'$-$\gamma$ transition form factors vs. $Q^2$. The solid lines represent the predictions obtained from the modified HSA using the AS wave function. Data are taken from PLUTO [5] (■), TPC/2$\gamma$ [6] (●) and CELLO [7] (○).

of the $\eta$ and $\eta'$. The PCAC result for the $\eta$ and $\eta'$ decays reads

$$\Gamma(\eta \to \gamma\gamma) = \frac{9\alpha^2 m_\eta^3}{16\pi^3} \left[\frac{C_8}{f_8}\cos\vartheta_P - \frac{C_1}{f_1}\sin\vartheta_P\right]^2$$
$$\Gamma(\eta' \to \gamma\gamma) = \frac{9\alpha^2 m_{\eta'}^3}{16\pi^3} \left[\frac{C_8}{f_8}\sin\vartheta_P + \frac{C_1}{f_1}\cos\vartheta_P\right]^2 \quad (10)$$

The three parameters, $f_1$, $f_8$ and $\vartheta_P$, are determined through a combined least square fit to the data on the form factors and the decay widths. The parameters acquire the following values:

$$f_1 = 145 \pm 3\,\text{MeV}, \qquad f_8 = 136 \pm 10\,\text{MeV}, \qquad \vartheta_P = -18° \pm 2° \quad (11)$$

($\chi^2 = 14.8$ for 18 data points). Since $f_1$ and $f_8$ have rather similar values nonet symmetry of the wave functions holds approximatively. The decay constants of the physical mesons are: $f_\eta = 175 \pm 10\,\text{MeV}$ and $f_{\eta'} = 95 \pm 6\,\text{MeV}$. The quality of the fit can be judged from Fig. 3 where fit and data[5,6,7] for the transition form factors are shown. As expected from the very good value of the $\chi^2$ the agreement between theory and experiment is excellent. The computed values for the decay widths are $\Gamma(\eta \to \gamma\gamma) = 0.50\,\text{keV}$ and $\Gamma(\eta' \to \gamma\gamma) = 4.17\,\text{keV}$. The value for the mixing angle is compatible with other results, see for instance, [15]. From chiral perturbation theory Gasser and Leutwyler [15] predicted a value of $170 \pm 7\,\text{MeV}$ for the $\eta_8$ decay constant whereas Donoghue et al. [16] found $163\,\text{MeV}$. In order to see whether or not such a large value is definitively excluded in the modified HSA the combined fit is repeated, keeping $f_8$ at the value of $163\,\text{MeV}$. The resulting fit is not as good as the precedent fit but still of acceptable quality ($\chi^2 = 20.7$). It provides: $f_1 = 143 \pm 3\,\text{MeV}$, $\vartheta_P = -21° \pm 1°$. The results for the form factors are almost as good as before. The resulting decay widths are $\Gamma(\eta \to \gamma\gamma) = 0.47\,\text{keV}$ and $\Gamma(\eta' \to \gamma\gamma) = 4.20\,\text{keV}$. In view of the experimental uncertainties in the $\eta \to \gamma\gamma$ decay width and of the moderate change of $\chi^2$ one cannot exclude the possibility of a $f_8$ as large as 163 MeV although a value around 140 MeV is favored. In contrast to $f_8$ the other two parameters, $f_1$ and $\vartheta_P$, are tightly constrained. The analysis of [4] provides no evidence for a sizeable gluon admixture to the $\eta_1$.

## 4. Summary

The AS wave function as the only phenomenological input leads to a good description of the $\pi \to \gamma$ transition form factor within the modified HSA. Results obtained with the frequently used CZ wave function are, on the other hand, in conflict with the data and should therefore be discarded. The use of the CZ wave function or other strongly end-point concentrated wave functions in the analyses of other exclusive reactions, e. g. $\gamma\gamma \to \pi\pi$ or $B \to \pi\pi$, seems unjustified and likely leads to overestimates of the perturbative contributions. The $\eta$-$\gamma$ and $\eta'$-$\gamma$ transition form factors are also analyzed within the modified HSA. Assuming for the SU(3) basis states again the AS wave function, the decay constants and the mixing angle are determined from a combined fit to the data on form factors and decay widths. The values found are in fair agreement with the results obtained in other analyses. One is tempted to conclude that the $\eta_8$ and $\eta_1$ wave functions are indeed correct approximately. One may also calculate the form factor for the transition of a virtual photon into a pion [17].